\documentclass[a4paper,onecolumn,preprint]{emulateapj}
\slugcomment{Accepted by ApJ}

\shorttitle{GRBs and SNe from collapsars}

\shortauthors{Lopez-Camara, Lee \& Ramirez-Ruiz}

\begin{document}

\title{GAMMA RAY BURSTS AND SUPERNOVA SIGNATURES IN SLOWLY ROTATING COLLAPSARS}

\author{Diego Lopez-Camara, William H. Lee} \affil{Instituto de
Astronom\'{\i}a, Universidad Nacional Aut\'{o}noma de M\'{e}xico,
Apdo. Postal 70-264, Cd. Universitaria, M\'{e}xico
D.F. 04510, (dlopez,wlee)@astroscu.unam.mx}

\and

\author{Enrico Ramirez-Ruiz} \affil{Department of Astronomy and
Astrophysics, UCO/Lick, University of California, Santa Cruz, Santa
Cruz, CA, 95060, enrico@ucolick.org}

\begin{abstract}
We consider accretion onto newborn black holes (BHs) following the collapse
of rotating massive stellar cores, at the threshold where a
centrifugally supported disk gives way to nearly radial inflow for low
angular momentum. For realistic initial conditions taken from
pre-supernova (pre-SN) evolution calculations, the densities and temperatures
involved require the use of a detailed equation of state and neutrino
cooling processes, as well as a qualitative consideration of the
effects of general relativity. Through two-dimensional dynamical
calculations we show how the energy release is affected by the
rotation rate and the strength of angular momentum transport, giving
rise to qualitatively different solutions in limits of high and low
angular momentum, each being capable of powering a gamma-ray burst (GRB). We explore the
likelihood of producing Fe-group elements in the two regimes and
suggest that while large and massive centrifugally supported disks are
capable of driving strong outflows with a possible SN-like signature,
quasi-radial flows lack such a feature and may produce a GRB without
such an accompanying feature, as seen in GRB060505.
\end{abstract}


\section{Introduction}\label{sec:intro}

Gamma-ray bursts (GRBs) have remained as an outstanding problem in
astrophysics since their discovery in 1967 \citep{kso73}.  Major
advances in their understanding came through observational
breakthroughs following BATSE, which provided an extensive all-sky
catalog \citep{fm95} and the identification of counterparts at lower
energies at high redshift \citep{m97} through Beppo Sax \citep{vkw00}
and Swift \citep{gcn07}. Isotropic equivalent energies range from
$10^{49}$~erg~s$^{-1}$ to $10^{52}$~erg~s$^{-1}$, coming from sources
at redshifts as high as $z\simeq 6$ that are randomly distributed over
the sky \citep{m92}. GRBs are commonly divided into short (SGRBs, less
than 2 seconds in duration, typically at redshift $z\simeq1$ and
frequently in host galaxies with low star formation rates (SFRs)) and long
(LGRBs, longer than 2 seconds, residing at $z\simeq1-5$ in star
forming galaxies) events. An additional distinction is that those of
the short variety tend to have harder spectra, hence the
short-hard/long-soft denomination \citep{d96}. Recently there appears
to be a substantial and, in some respects, puzzling overlap between
the two populations, and it has been suggested that a third group or
even a new classification is necessary \citep{g06,gy06,dv06nat,f06}.

The energy release, short duration, and variability time scales
associated with GRBs favor accretion onto compact objects, or
magnetically powered events from such objects, as their ultimate
source. Many models have been suggested over the years, a number of
which were discarded once the global energy budget was fixed through
the resolution of the distance scale to the sources. Among the early
proposals still considered are compact binary mergers
\citep{ls74,ls76,bp86,bp91,e89,n92}, magnetars \citep{u92,t94,mr97}
and the collapse of massive stellar cores \citep{w93}, see
\citep{m02,piran04,n07,lrr07} for reviews.

In general, accretion onto a central object can proceed and release
gravitational binding energy efficiently if the gas can cool or
otherwise lose its internal energy e.g., through advection, and thus
move down in the potential well \citep{s64,z64}. At the usual
astrophysical scales, this occurs through photons, and produces
phenomena ranging from active galactic nuclei (AGN) to low mass X-ray
binaries (LMXBs). If the accretion rate is too high though, the fluid
may become so optically thick to the photons that they become trapped
and are unable to cool the gas. At even higher accretion rates, a
different mechanism enters the picture, namely, neutrino cooling. This
regime is termed ``hypercritical", since it is far above the usual
Eddington limit for photons, and is relevant, for example, in post-supernova cores and the associated fall back, as was probably the case
in SN1987A \citep{c89,hc91}. It is this phenomenon that is now thought
possible of generically powering GRBs. The question at the forefront
of attention has been how to achieve these conditions in an
astrophysical setting.

There has been mounting evidence in the past few years linking LGRBs
at low redshift with type Ic SNe, with spatial and temporal
coincidences for a number of events \citep{wb06,kaneko08}. \citet{g98}
linked for the first time a LGRB with a SN (GRB980425 with
SN1998bw). After 27 days the optical spectrum of GRB980425 resembled
that of a H-deficient type Ic SN. To date there have been at least 6
other LGRB/SN connections: XRF020903 \citep{sod05}; GRB021211 with
SN2002lt \citep{dv03}; GRB030329 with SN2003dh (whose optical spectrum
afterglow was strikingly similar to that of SN1998bw \citep{s03});
GRB031203 with SN2003lw \citep{m04,rr05}; GRB050525A with SN2005nc
\citep{dv06apj}; and GRB060218 with SN2006aj \citep{c06}. In addition,
LGRBs in general are associated with low metallicity star-forming
regions in their host galaxies \citep{p04,g05,sol05,fruchter06,f06}
suggesting a link to massive stars \citep{bkd02,wb06}.

The observational association with SNe has given support to the idea
that LGRBs are produced in the collapse of massive stellar
cores. \citet{mw99} developed this in great detail in the collapsar
scenario, assuming a black hole (BH) is formed at the center and evaluating
the potential to produce a GRB. The infalling stellar material that is
not ejected in the usual SN forms a massive, dense and hot
accretion disk which releases the necessary energy to power the
burst. They distinguished between Type I and Type II collapsars: the
former entails the direct collapse of the iron core to a BH,
while in the latter initially a proto-neutron star lies at the center
of the star, and later collapses to a BH once it has accreted
a sufficient amount of mass. A GRB may be produced even if there is no
BH at the center in a magnetically dominated explosion during
the proto-neutron star phase of the collapse \citep{dessart08}. Here,
we will focus on Type I collapsars. The absence of H lines in the
observed GRB/SN spectra implies that the progenitor has somehow lost
its envelope, perhaps through interaction with a binary companion
and/or strong winds, and is thus essentially a Wolf-Rayet (WR) star
\citep{izzard04,cantiello07}.

In the standard version of the collapsar, the GRB/SN link is a quite
natural consequence, and observed associations have thus provided a
strong motivation in this sense to study it further. Typical LGRBs
have $z\simeq1-4$, which is too far for a SN counterpart to be
detected, and indeed, the cases in which there is such a
correspondence all lie at fairly low redshift. Two recent events
however, GRB060505 and GRB 060614 \citep{f06,g06,gy06,dv06nat}, have
pointed perhaps to a different situation in which the GRB is {\em not} accompanied by a SN. Their low redshift
allowed deep searches which convincingly ruled out the presence of an
underlying type Ic stellar explosion of the kind seen in SN1998bw and
SN2003dh, and particularly, the host of GRB060505 is a star-forming
galaxy, similar to that of typical LGRBs \citep{f06}. We explore here
how the initial angular momentum distribution in the star, and
possibly the vigor of angular momentum transport within the
centrifugally supported accretion disk, may furnish a key ingredient
to understand this behavior.

An important point is that despite considerable effort (see \citet{wh06}), the stellar rotation rate in pre-SN cores is
not fully determined. It can sensitively depend on evolutionary
details, such as mass loss on the main sequence and the stellar
magnetic field, which both lead to important spin down in later stages
\citep{spruit02, hws05}. Binary interactions may also affect the
rotation rate through tidal interactions \citep{detmers08}. We note
that \citet{yoon05}, and \citet{wh06}, have identified a channel for
massive stars in which mass and angular momentum losses are greatly
reduced by complete mixing in the main sequence and the consequent
absence of a giant phase.

What we do know is that the specific angular momentum, $J$, needed to
form a disk around the BH is at least the value of the innermost
stable circular orbit $R_{isco}$ ($R_{isco}=3r_{\rm g}$, where
$r_{\rm g}=2GM_{\rm BH}/c^{2}$ is the Schwarzschild radius) times the
velocity of the particles (which are very close to the speed of
light). So, a first estimate for the critical J would be $J_{crit}=R_{isco}c=3r_{\rm g}c=6GM/c\sim10^{16}$~cm$^{2}$~s$^{-1}$ for
a one solar mass BH (the true critical value is $J_{\rm crit}\simeq2r_{\rm g} c$, see Sextion~3.1).  Most studies of collapsars and neutrino-dominated accretion flows
\citep{mw99,pwf99,h00,npk01,pb03,pmab03,lrrp05} have considered
angular momentum distributions that are well above this limit and
essentially guarantee the formation of a centrifugally supported
accretion disk (in stellar core collapse calculations as well as disk
evolution studies). In practice, there is also something akin to a
maximum value of angular momentum for the collapsar model to work,
since for very rapid rotation the accretion disk forms at large radii
and the binding energy cannot be effectively dissipated as neutrinos
\citep{mw99,lrr06}.  Clearly in the limit of slow rotation the
accretion disk will form very near the BH, and so general relativity
(GR) will play an important role in its evolution.

Our general objective is to explore the morphology, energy release, and
observable signatures of hypercritical accretion flows from massive
stellar core collapse for a range of angular momentum values covering
the transition from centrifugally supported disks to near radial
inflow (lying below those usually considered in collapsar
calculations) in order to better understand the possible production of
LGRBs as well as their putative progenitors. We improve upon previous
studies \citep{mw99,pmab03,lrr06} in significant ways by considering
more detailed thermodynamics in the equation of state, an improved
treatment of neutrinos, and through realistic initial conditions taken
from evolutionary calculations of \citet{wh06}. 

We first describe the input physics and numerical setup in
Section~\ref{sec:input}, followed by our results in
Section~\ref{sec:results}. Prospects for GRB production and the link
between SN and GRBs are given in Section~\ref{sec:conc}. \\

\section{Input physics}\label{sec:input}
 
\subsection{Equation of State and cooling processes}\label{sec:eos}

We use a detailed equation of state \citep{lrrp05} where the total
pressure, $P$, contains contributions from an ideal gas of $\alpha$
particles and free nucleons in nuclear statistical equilibrium (NSE),
$P_{\rm gas}$, blackbody radiation, $P_{\rm rad}$ (the optical depth
to photons in the gas is such that they are fully trapped), neutrinos,
$P_{\nu}$, and relativistic electron/positron pairs of arbitrary
degeneracy, $P_{e^{\pm}}$ \citep{bdbn96}. We allow for neutronization
and a correspondingly variable electron fraction $Y_{\rm e}$ by
requiring charge neutrality and equilibrium in weak interactions,
depending on whether the fluid is optically thick or thin to its own
neutrino emission \citep{b03,lrrp05}.

The physical conditions in the accretion flow within the collapsing
stellar core are such that the temperature and density is $T\simeq10^{9}-10^{10}$~K, and $\rho\simeq10^{8}-10^{10}$~g~cm$^{-3}$
respectively. Thus the reaction rates that will dominate the neutrino
emissivities, $\dot{q}_{\nu}$, are $e^{\pm}$ capture onto free
nucleons ($\dot{q}_{cap}$) and $e^{\pm}$ annihilation
($\dot{q}_{ann}$), for which we use the tables of \citet{lmp01} and
the fitting functions of \citet{i96}, respectively \footnote{Neutrino emission
through plasmon decays and nucleon$-$nucleon bremsstrahlung was also
computed and found to be insignificant compared to capture and
annihilation.}. Finally, photodisintegration and synthesis of $\alpha$
particles can also cool or heat the gas, and is correspondingly
accounted for (see also \citet{lrrp05}).

If the fluid is optically thin, the total cooling is simply the sum of
the emissivities described above. At a finite optical depth, we may
split it into scattering and absorption components as $\tau_{\nu} =
\tau_{\rm scat} + \tau_{ \rm abs}$, where the absorption term is due
to the inverse reaction of $e^{\pm}$ capture onto protons or neutrons,
$\tau_{\rm abs-cap}$, and $\nu \overline{\nu}$ annihilation,
$\tau_{\rm abs-ann}$ \citep{dmpn02}:
\begin{equation} \label{tau} 
\tau_{\rm abs} = \tau_{\rm abs-cap} + \tau_{\rm abs-ann}, 
\end{equation}
with
\begin{eqnarray}
\tau_{\rm abs-cap} & = & \frac{2 \ \dot{q}_{cap} \ H}{7 \ \sigma_{\rm SB} \
T^{4}} \\ 
\tau_{\rm abs-ann} & = & \frac{2 \
\dot{q}_{ann} \ H}{7 \ \sigma_{\rm SB} \ T^{4}},
\end{eqnarray}
$H$ being the pressure scale height in the disk and $\sigma_{\rm SB}$
the Stefan$-$Boltzmann constant.

The contribution from scattering off free nucleons is
\begin{equation}
\tau_{\rm scat} = 13.8 \left( C_{s,p} Y_{p} \ + \ C_{s,n} Y_{n}
\right) \frac{\sigma}{m_{\rm u}} \left(k_{\rm B} T / m_{e}c^{2}
\right)^{2} \rho H,
\end{equation}
where $Y_{n}$ and $Y_{p}$ are the neutron and proton fractions,
$C_{s,p}=(1+5\alpha^{2})/24$, $C_{s,n}=[4(C_{v}-1)^{2} +
  5\alpha^{2}]/24$, $C_{v}=0.5+2\sin^{2}\theta_{w}$,
$\sin^{2}\theta_{w} \approx 0.23$, $\sigma=1.76 \times
10^{-44}$~cm$^{2}$, and $\alpha=1.25$ \citep{st83} and the rest of the
symbols have their usual meanings.

To compute the internal energy, cooling rate, and pressure due to
neutrinos we used a two-stream approximation
\citep{pn95,dmpn02,jypdm07}:
\begin{equation}
E_{\nu} \left( \tau \right) = 3 P_{\nu} \left( \tau \right) 
= \frac{7
\pi^{2} (k_{B} T)^{4}}{8 \cdot 15 (\hbar c)^{3}} \left(
\frac{(\tau_{\rm abs} + \tau_{\rm scat})/2 + 1/\sqrt{3} }{(\tau_{\rm
abs} + \tau_{\rm scat})/2 + 1/\sqrt{3} + 1/3\tau_{\rm abs}} \right) \
\ \ \mbox{erg cm$^{-3}$ },
\end{equation}
\begin{equation} \label{qnu}
\dot{q}_{\nu} \left( \tau \right) = \frac{7}{8} \left( \frac{4
\sigma_{SB} T^{4}/3}{ (\tau_{\rm abs} + \tau_{\rm scat})/2 +
1/\sqrt{3} + 1/3\tau_{\rm abs}}\right)\left(\frac{H}{\rm cm}\right) \
\ \ \mbox{erg cm$^{-3}$ s$^{-1}$},
\end{equation}
and
\begin{equation} 
P_{\nu} \left( \tau \right) = \frac{1}{3} \left( \frac{7 \pi^{2}
(k_{B} T)^{4}}{8 \cdot 15 (\hbar c)^{3}}\right) \left(\frac{
(\tau_{\rm abs} + \tau_{\rm scat})/2 + 1/\sqrt{3}}{ (\tau_{\rm abs} +
\tau_{\rm scat})/2 + 1/\sqrt{3} + 1/3\tau_{\rm abs}} \right).
\end{equation}
The total neutrino luminosity (in erg~s$^{-1}$) is then calculated
through
\begin{equation} \label{lnu}
L(\tau)= \int \dot{q}_{\nu} \left( \tau \right) \ dV \ \ \ \mbox{erg \
s$^{-1}$}.
\end{equation}

\begin{figure}
  \begin{center}
    \includegraphics[width=0.8\textwidth]{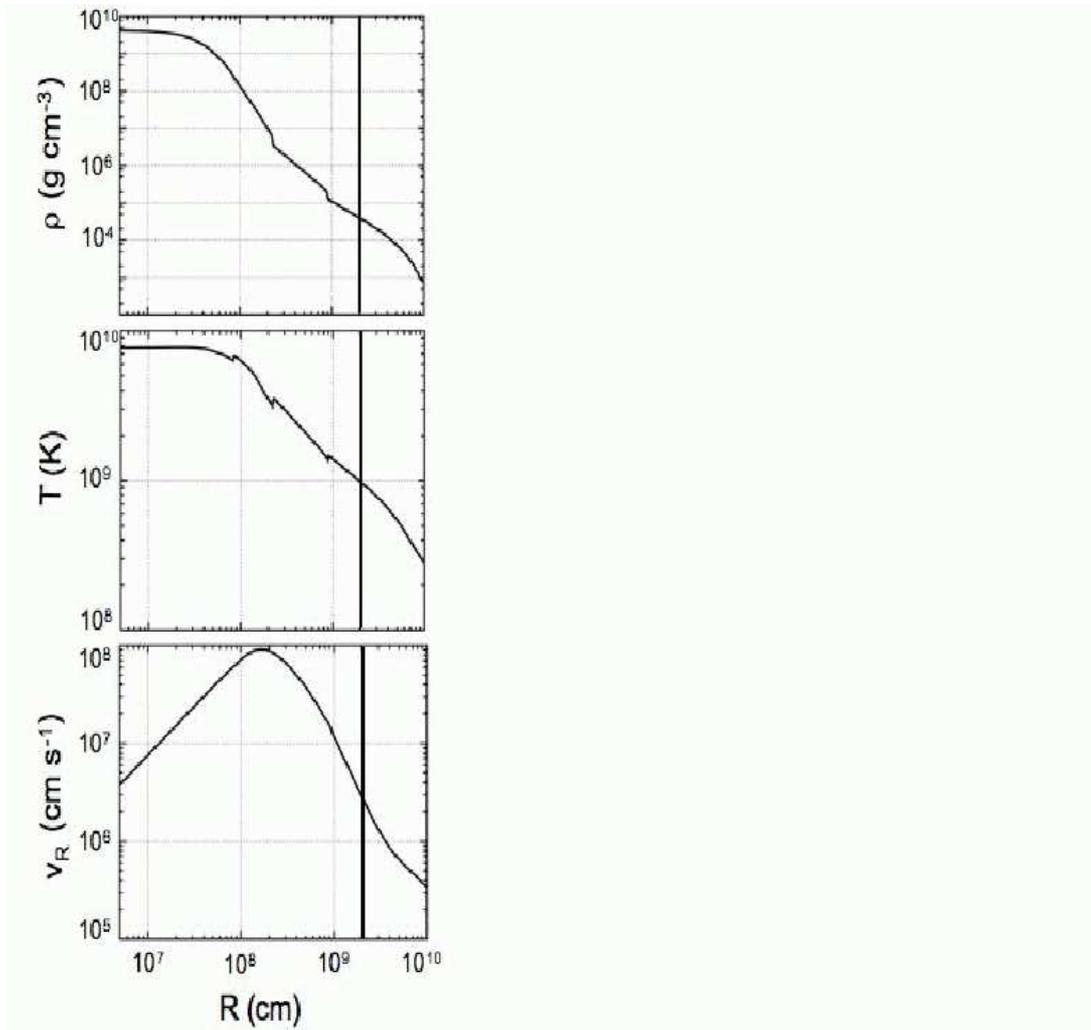}
    \caption{Density, temperature and radial velocity profiles in the
      pre-SN star used as an initial condition (model 16TI of
      \citet{wh06}). The solid vertical line in each panel marks the
      outer computational boundary used in the simulations.}
    \label{IC}
  \end{center}
\end{figure} 
 
\subsection{Initial setup and numerical method}\label{sec:ICs}

Our initial data were taken from the one-dimensional pre-SN
calculations of \citet{wh06}. Specifically we considered model 16TI, a
rapidly rotating ($v_{\rm rot}=390$~km~s$^{-1}$ at the equator),
16$\,M_\odot$ WR star of low metallicity (1\% solar), low mass
loss (2$\,M_\odot$ loss at the end of the evolution), with an iron
core surrounded by silicon and oxygen, neon and carbon shells. The
density, temperature, and radial velocity distributions as functions of
the spherical radius, $R$, were mapped onto a two-dimensional configuration
assuming spherical symmetry. We further suppose that the low angular
momentum iron core (with mass $M_{Fe}=1.6M_{\odot}$) will promptly
collapse to a BH (possibly producing a Type~I collapsar, see
Section~\ref{sec:intro}) and thus condense all of this matter onto the
origin in a point mass at the start of the simulation, when the
surrounding envelope begins its infall.

The calculations are performed with a two-dimensional Smooth Particle
Hydrodynamics (SPH) code \citep{mon92} in cylindrical coordinates
$(r,z)$ with azimuthal symmetry \citep{lrr02,lrr06,lrrp05} and a
reflecting boundary along the polar axis (no reflection about the
$z=0$ plane is assumed). The inner and outer boundaries were set at
spherical radii $R_{\rm in} = 2\times 10^{6}$~cm and $R_{\rm out} = 2
\times 10^{9}$~cm, respectively, with free outflow conditions imposed
during the simulation. Any material crossing the inner boundary is
assumed to be accreted by the BH, whose mass is
correspondingly updated. Matter lost through the outer boundary is not
followed. We solved the continuity, momentum, and energy equations,
including the terms arising from the full viscous stress tensor $t_{r
  \phi}$ \citep{lrr02}, and took the $\alpha$ prescription
\citep{ss73} for the coefficient of viscosity, $\eta_{\rm v}=\alpha
\rho c_{\rm s} \ H$, where $c_{\rm s}$ is the local sound speed and
$H= c_{\rm s} / \Omega_{K}$ is the pressure scale height ($\Omega_{K}$
is the local Keplerian orbital angular frequency). Doing the
calculation in two dimensions allows for good spatial resolution, a
considerable simulated interval and a solid discussion of angular
momentum and viscosity effects. The initial profile is reproduced with
a Monte Carlo accept/reject procedure, typically with $1-2 \times
10^5$ SPH particles in the computational domain. The spatial
resolution is intrinsically adaptive in this numerical scheme and
varies greatly, increasing at small radii (or equivalently, high
densities) with a typical smoothing length of few $\times 100$~m in
the inner disks formed.

The gravitational potential of the BH is computed with the
pseudo-Newtonian expression of \citet{pw80}:
\begin{equation}
\phi = - \frac{G M_{\rm BH}}{r - r_{\rm g}},
\end{equation}
which approximates GR effects in the inner regions for a nonrotating
(Schwarzschild) BH. In particular, it reproduces the position
of the innermost stable circular orbit at $r_{isco}=3r_{\rm g}$, as
well as the marginally bound orbit at $r_{mb}=2r_{\rm g}$.

The mass in the collapsing envelope is comparable to that of the BH, so one should consider self-gravity. We assume that the mass
distribution remains roughly spherical, so a mass element at spherical
radius $R_{0}$ is affected only by the matter distribution at radii $R
\le R_{0}$ as if it were concentrated at the center of the star. This,
while strictly valid only for configurations with spherical symmetry,
is nevertheless a good approximation for the present set of
calculations, since deviations in the mass distribution indeed remain
small. Figure~\ref{IC} shows the initial density, temperature, and
radial velocity profiles in the pre-SN star as functions of
spherical radius, taken from \citet{wh06}.

\section{Results}\label{sec:results}

\subsection{Expectations}\label{sec:exp}

Consider a nonrotating BH surrounded by rotating material
distributed in spherical symmetry with vanishing pressure. For a
distribution of specific angular momentum increasing monotonically
with respect to the polar angle, flowlines in the polar regions will
be qualitatively different than those near the equator, and can be
divided into three types: (1) those with very little angular momentum,
moving nearly radially into the BH; (2) those with high
angular momentum which can find a point where centrifugal support
balances the gravitational field (the circularization radius); and (3)
those that would accrete onto the central mass but cross the
equatorial plane, $z=0$, before doing so. Recall that in GR capture
orbits exist even for nonvanishing angular momentum, and the critical
value defining this class of solutions is $J_{\rm crit} \simeq
2.0r_{\rm g} c$. Here we will consider $J \ge J_{\rm crit}$ as high
angular momentum, and $J \sim J_{\rm crit}$ as low. This general
situation has been considered before in the context of LMXBs by \citet{bel01}, and for accretion modes in collapsars by \citet{lrr06} (an analytical derivation of low angular momentum flow lines in GR has been carried out by \citet{mtn09}).

Encountering the centrifugal barrier or a flowline from the opposite
hemisphere can lead to a shock (and thus hydrodynamical effects) in
which the kinetic energy is efficiently converted into thermal
energy. If the material does not have enough angular momentum to
remain in orbit, it will fall onto the BH due to GR effects,
{\em even in the absence of any angular momentum transport mechanism}
through a fast, nearly {\em inviscid} disk. For high angular momentum,
a disk (with scale height $H$ and typical radius $r$) forms in the
equatorial plane, and if cooling occurs some of this energy will be
lost from the system through photons or neutrinos, depending on the
physical conditions.  For efficient cooling the disk will be nearly
isothermal and geometrically thin, with $H\ll r$. In the opposite
adiabatic regime it can be quite thick, and $H \sim r$. Energy
dissipation will in general circularize motions at a radius given by
the local value of angular momentum, and if any of this is
subsequently removed (through any mechanism), or it is already below
the critical value, the material inside the disk may move radially and
possibly be accreted by the BH. The size and geometry of the
newly formed disk thus reflects: (1) the angular momentum distribution,
fixing the point where the centrifugal barrier is encountered; (2) the
mass accretion rate, giving the total gravitational energy which can
be converted into thermal energy; (3) the cooling rate, determining
the disk geometry; (4) the value of the $\alpha$ parameter,
responsible for angular momentum transport. 

In general, previous collapsar studies or neutrino-dominated accretion
flow studies have considered cases where $J \gg J_{\rm crit}$
\citep{mw99,pwf99,h00,npk01,pb03,pmab03}. Only recently has the low
angular momentum regime in the collapsar context been addressed
\citep{lrr06}, showing the presence of the small {\em dwarf} disk as
described above. For $J > 2.0r_{\rm g}c$, a thick disk forms, supplying
the bulk of the neutrino luminosity from a shocked toroidal region
around the BH. More recently, \citet{jp08} considered the
minimum angular momentum required for disk formation, taking into
account that the accretion of mass by the central object raises the
angular momentum threshold as the stellar core collapses.

We explored the dynamical evolution for different values of the
$\alpha$ viscosity parameter (see Table~1), and more importantly,
different values of angular momentum, which were always separated into
radial, $R$, and polar angle, $\theta$ (measured form the rotation
axis), variations as $J=J(R,\theta)=J_{\rm R}(R)
\ J_{\theta}(\theta)$.  For the radial component, $J_{\rm R}(R)$, we
considered constant values with radius (see Table~1). We will consider
more realistic distributions of $J_{\rm R}(R)$ in future work
(D. Lopez-Camara et al. 2008, in preparation). In all cases the initial
distribution for $J(\theta)$ corresponded to rigid body rotation on
shells, $J(\theta)= \sin^{2}\theta$.

\begin{table}[!h]\centering
\caption{Model Parameters}
\begin{tabular}{cccc}
\hline
\hline
$J_{0} $ & $\alpha$ \\
& ($r_{\rm g}c$) &  \\
\hline
2.0 & 0.10 \\
2.5 & 0.10 \\
3.0 & 0.10 \\
\hline
2.0 & 0.01 \\
2.5 & 0.01 \\
3.0 & 0.01 \\
\hline
2.0 & 0.00 \\
2.5 & 0.00 \\
3.0 & 0.00 \\
\hline
\hline
\end{tabular}
\end{table}

\subsection{Global properties and flow morphology}\label{Jcte}

\begin{figure}
  \begin{center}
    \includegraphics[width=0.6\textwidth]{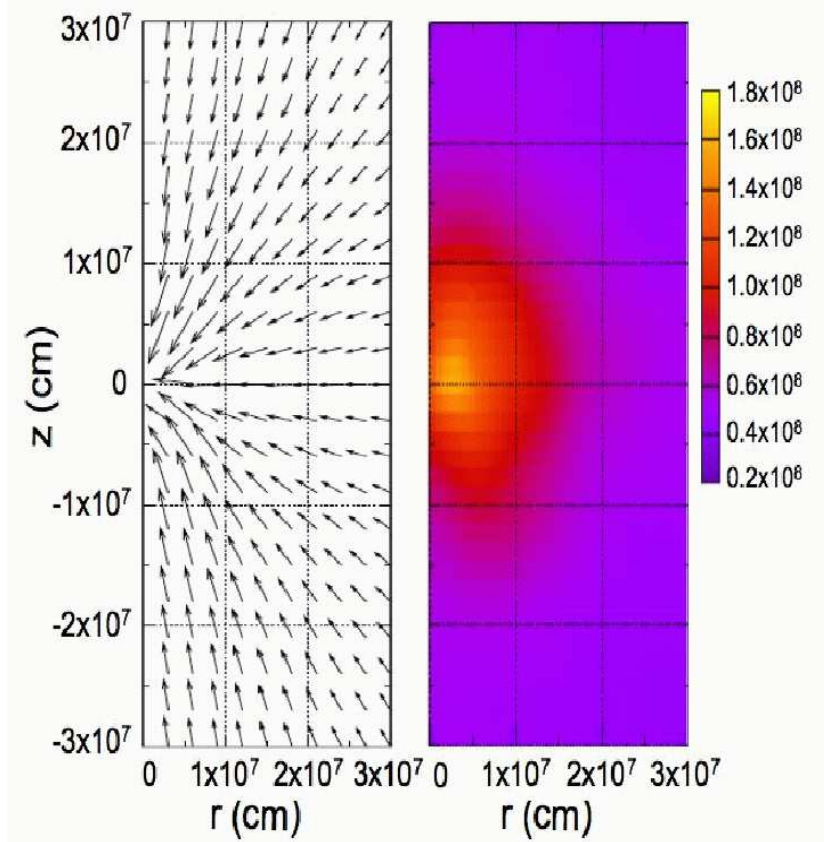}
    \caption{Velocity field and density map (in g cm$^{-3}$) for
      $(J_0,\alpha)=(2r_{\rm g}c,0.1)$ at $t=0.2$~s. The largest
      vector corresponds to $v\approx8\times10^{7}$ cm s$^{-1}$.}
    \label{vel1}
  \end{center}
 \end{figure}

\begin{figure}[!h]
  \begin{center}
    \includegraphics[width=0.7\textwidth]{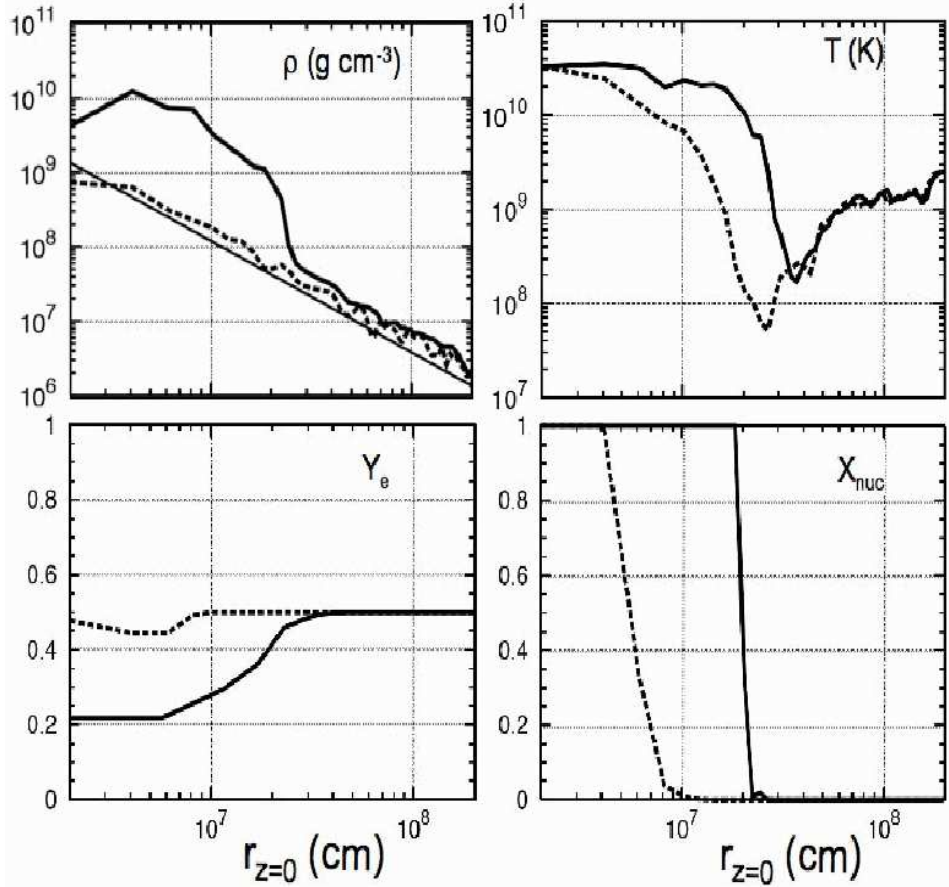}
    \caption{Equatorial ($z=0$) density, temperature, electron
      fraction and nucleon mass fraction distribution for
      $(J_0,\alpha)=(3 r_{\rm g}c,0.1)$ (solid line) and
      $(J_0,\alpha)=(2 r_{\rm g}c,0.1)$ (dashed line) at
      $t=0.2$~s. The power law given by a thin solid line in the
      density panel is the solution given in Equation~(10).}
    \label{rho4z}
  \end{center}
\end{figure}

For $J(R) = 2.0 r_{\rm g}c$, the flow is essentially at the critical
value, and remains nearly radial. Even though some compression occurs
in the equatorial region the centrifugal barrier is absent
and no shocks are formed.  Moreover, since $J(R)$ is constant, the
initial velocity field and density map (shown in Figure~\ref{vel1}) do
not vary substantially as the simulation progresses. For an inflow in
strict free fall, conservation of mass and energy give a density profile 
\begin{equation}
\rho=\frac{\dot{M}}{4 \pi r^{2}v_{r}}=1.2 \times 10^{8} \left(
\frac{\dot{M}}{\mbox{0.5 M$_{\odot}$~s$^{-1}$}}\right)
\left(\frac{r}{10^{7}{\rm cm}} \right)^{-3/2} \left( \frac{M_{\rm BH}}{1.7
  M_{\odot}}\right)^{-1/2} \mbox{g~cm$^{-3}$},
\end{equation}
where $\dot{M}$ is the mass accretion rate and $M_{\rm BH}$ the BH mass.  The solution is plotted as a dotted line in
Figure~\ref{rho4z} with the above scalings set to unity, clearly
showing the excellent agreement between the full simulation and this
simple estimate along the equator. The energy release, however, is not
negligible, as will be seen below in Section~\ref{mdot_eff_lum}. As long as
the equatorial angular momentum is below this critical value, the
solution is insensitive to the actual value of the viscosity
parameter, $\alpha$, that is chosen (we computed solutions for
$\alpha$ in the range $0-10^{-1}$).

\begin{figure}  
  \begin{center}
    \includegraphics[width=0.6\textwidth]{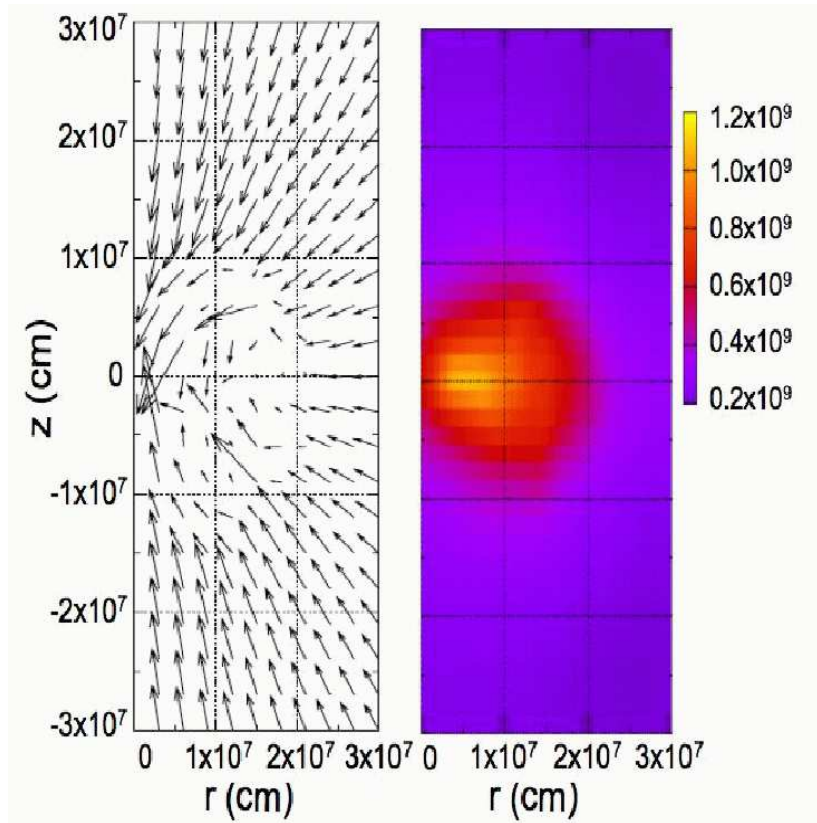}
     \caption{Velocity field and density map (in g cm$^{-3}$) for
       $(J_0,\alpha)=(3 r_{\rm g}c,0.1)$ at $t=0.2$~s. The largest
       vector corresponds to $v\approx8 \times10^{7}$ cm s$^{-1}$ for
       densities lower than $5 \times 10^{8}$~g~cm$^{-3}$. At higher
       values the velocity vectors are reduced by a factor of two. The
       location of the accretion shock is clearly seen at $r(z=0)
       \approx 2 \times 10^{7}$~cm. }
    \label{vel2}
  \end{center}
\end{figure}

For values of $J(R)$ above the critical value the outcome changes
drastically. The scale of the disk correlates with $J_{0}$ so we will
limit the discussion to the case with $J(R) = 3.0r_{\rm g}c$ and
$\alpha=0.10$. An initial disk forms close to the equator after 0.1~s,
the free-fall time from the boundary of the Fe core, when the
innermost regions of the envelope approach their circularization
radius $r_{\rm c}\approx J_{0}^{2}/GM_{\rm BH}$. The transformation of
kinetic into thermal energy at the centrifugal barrier produces a hot
torus around the BH. Figure~\ref{vel2} shows the velocity
field and a density map at $t=0.2$~s for $(J_0,\alpha)=(3 r_{\rm
  g}c,0.1)$. Note how in the velocity field the flow lines in the
polar regions are still largely radial, and an asymmetry is apparent,
with the polar shock lying substantially closer to the black hole than
in the equator. A large-scale meridional circulation with two
prominent eddies is also apparent.

The temperature and density rise rapidly behind the shock front,
releasing a substantial amount of energy and photodisintegrating
He. It is here where neutrino emission, finite optical depth effects
and neutronization can become important.  A substantial fraction of
the released energy will likely be directed to the polar regions
because of geometrical effects, and may give rise to a fireball
capable of producing a GRB (see Section~\ref{mdot_eff_lum}).

Outside the shock front, the solution is essentially the same as for
low angular momentum (see Figure~\ref{rho4z}), with a jump at
$r\approx r_{\rm c}$. For angular momentum $J_{0}=A r_{\rm g}c$, with
$A \geq 2$ being a constant, the circularization radius for equatorial
matter is $r_{\rm c}\approx 2A^{2} r_{\rm g}$ and the free-fall
velocity at $r_{\rm c}$ is $v_{\rm ff}=c/2^{1/2}A$. If the kinetic
energy of infall is entirely transformed into thermal energy, one
would naively estimate the latter as
\begin{equation}
kT=\frac{m_{\rm p} c^{2}}{12 A^{2}}\approx 10 (A/3)^{-2} {\rm MeV}.
\end{equation}
This is clearly an overestimate of the temperature, since the velocity
is not entirely radial but has a substantial azimuthal component,
$v_{\phi}$, by the time it reaches the circularization radius,
$r_{\rm c}$. It does nevertheless give a useful estimate (and upper
bound) on the postshock temperature, as can be seen from
Figure~\ref{rho4z}. As we shall see, it is also high enough to
guarantee the complete photodisintegration of the infalling nuclei
into their constituent neutrons and protons.

In the initially small disk that forms at the centrifugal barrier, if
$\alpha \neq 0$ transport processes transform the constant
distribution of angular momentum into a nearly Keplerian one where
differential rotation is important. For the Newtonian case the orbital
frequency is $\Omega \propto r^{-3/2}$ and at a few gravitational
radii this is not a bad approximation even in the pseudo-GR PW
potential. In a way this partially erases the memory of the initial
distribution: as long as the rotation rate is above a certain
threshold, the inner disk will quickly converge to a centrifugally
supported structure with strong differential rotation, independently
of the details of the initial angular momentum profile. The associated
shear produces dissipation and pumps mechanical energy into thermal
energy. The rotational structure of the inner disk depends on whether
$\alpha$ is finite or not. If $\alpha \neq 0$ the actual
value of the specific angular momentum can grow beyond the initial
equatorial value $J_{0}$ through transport. For the inviscid regime,
$\Omega \propto r^{-3/2}$ (and hence $J \propto r^{1/2}$) only as long
as this implies $J \leq J_{0}$. At greater radii the equatorial flow
maintains constant specific angular momentum. 

Approximately 0.2~seconds after the formation of the initial disk, a
shock begins to propagate outward, initially moving at $v\approx 2
\times10^{8}$~cm~s$^{-1}$ in the equatorial regions. The postshock
gas pushes out against the infalling envelope mainly because of
viscous heating for $\alpha=0.1$, and is aided by a combination of
neutrino heating and He synthesis \footnote{In a test calculation with the
viscous heating terms switched off, the shock had not started to move
outward after 0.5 seconds.}. The calculations were stopped at $t\sim
0.5$~s, when the shock reached the outer boundary. Not only is the
numerical solution no longer self-consistent, but by then our
resolution had decreased notably, since accretion onto the BH entails a loss of particles, and thus resolution. We will address this
issue in an adaptive form in future work. For low viscosity the inner
disk is denser due to the reduced transport efficiency and the outward
motion of the shock is only slightly delayed. In the inviscid limit
there is no associated heating but the piling up of material due to
the lack of transport also induces outward motion of the shock front
after a comparable delay of $\simeq 0.2$~s.

In general, the inner disk can remain thick despite the presence of
cooling because of the continous infall of material and, in some
cases, because the optical depth to neutrinos is large enough to
keep the internal energy from immediately escaping. As long as the
angular momentum of the infalling gas is greater than the critical
value, the morphology is similar, with a greater radial extent for
higher $J_{0}$.

The breaking of spherical symmetry is in principle a problem for the
computation of self-gravity, described in Section~\ref{sec:ICs}. We thus
checked how far the actual flow departs from spherical symmetry in
terms of the mass distribution as a function of the polar angle
$\theta$, and find that the scatter in the radially integrated mass
along cones with constant $\theta$ is small enough (10\%) to be
ignored in a first treatment.

\begin{figure}[!h]
  \begin{center}
    \includegraphics[width=0.6\textwidth]{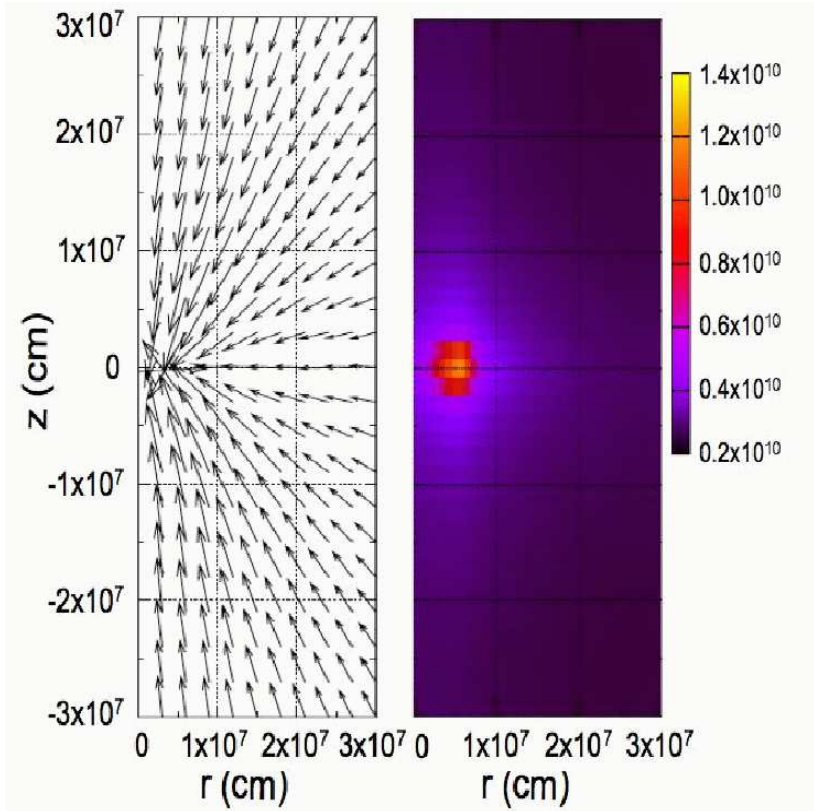}
     \caption{Velocity field and density map (in g cm$^{-3}$) for
       isothermal flow with $J_0=3r_{\rm g}c$ and $\alpha=0.1$ at
       $t=0.2$~s.}
    \label{isoflow}
  \end{center}
\end{figure}

\subsection{The importance of cooling}\label{coolingexple}

\begin{figure}  
  \begin{center}
    \includegraphics[width=0.6\textwidth]{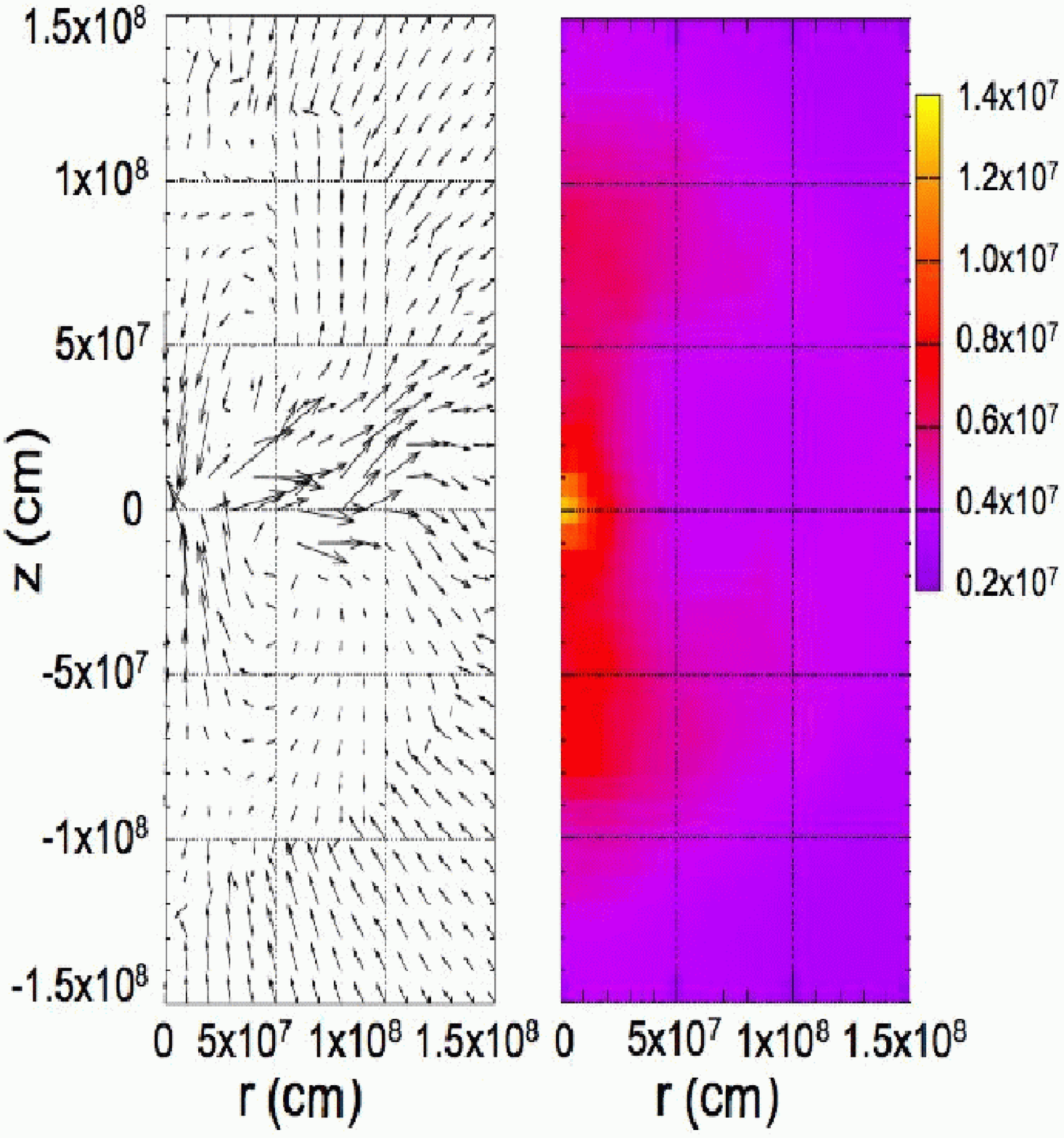}
     \caption{Velocity field and density map (in g cm$^{-3}$) for
       adiabatic flow with $J_0=3r_{\rm g}c$ and $\alpha=0.1$ at
       $t=0.2$~s. Note the different spatial scale compared to the
       previous figures. The largest velocity vectors correspond to
       $10^{8}$cm~s$^{-1}$ for densities lower than $5 \times
       10^{7}$g~cm$^{-3}$. At higher values the vectors are reduced by
       a factor of 2.}
    \label{adflow}
  \end{center}
\end{figure}

\begin{figure}
  \begin{center}
    \includegraphics[width=0.6\textwidth]{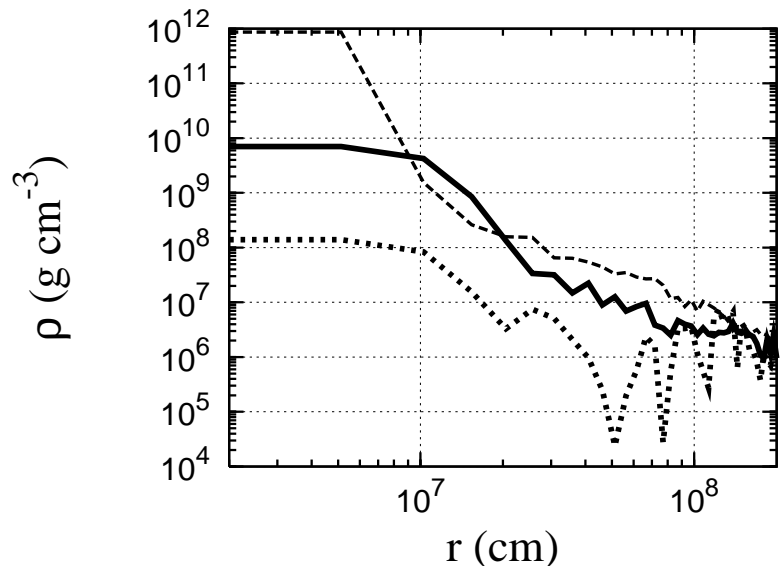}
    \caption{Equatorial ($z=0$) density distributions at $t=0.2$~s for
      the simulations carried out in the adiabatic (dotted) and
      isothermal (dashed) limits, both for $(J_0,\alpha)=(3 r_{\rm
        g}c,0.1)$. The thick solid line is the simulation with full
      microphysical treatment and $(J_0,\alpha)=(3 r_{\rm g}c,0.1)$.}
    \label{isoadfull}
  \end{center}
\end{figure}

To highlight and better understand the importance that the proper
computation of cooling has on the global properties of the solution,
we have calculated the evolution of the flow in two simplified cases.
Since neutrinos are the only means other than advection onto the BH through which the cas can cool, and thus move lower in the
gravitational potential well, there are two limits in this respect:
adiabatic inflow, in which no cooling occurs, and isothermal flow, in
which on the contrary, it is extremely efficient. The true solution
must lie somewhere between these two extremes, and it is instructive
to know which it resembles the most. We computed these, in one case by
eliminating the cooling terms in the energy equation and thus impeding
the outward flow of energy through neutrinos, and in the other by
using and ideal gas equation of state with $P= (\gamma-1)\rho u$,
where $\gamma=1.01$ and the same initial conditions, thus mimicking
the isothermal case where $\gamma=1$ and compressibility is very
high. The velocity fields and the corresponding density maps (also for
$J_0=3 r_{\rm g} c$ and $\alpha=0.1$ are shown in
Figures~\ref{isoflow} and \ref{adflow}. Equatorial density profiles
for the various cases are shown in Figure~\ref{isoadfull}, and can be
compared with those in Figure~\ref{vel1}).The isothermal flow looks
qualitatively similar to the low angular momentum case, but the
density is 2 orders of magnitude larger. There is little temporal
evolution once the centrifugal barrier is reached, and the solution is
quasi-stationary. On the other hand, in the adiabatic solution as soon
as the infalling gas reaches the centrifugal barrier, the shock
bounces rapidly outward and the subsequent expansion produces a
strong flow reversal, sweeping the inner envelope outward and beyond
the outer boundary (compare with Figure~\ref{vel2} and note that the
spatial scale is 5 times larger). At the instant shown in
Figure~\ref{isoadfull} the density has decreased considerably, even
beyond the solution where neutrino emission was included. The fact
that the accretion disk remains geometrically thick in the calculation
with neutrino cooling indicates that it is not extremely efficient,
resembling the adiabatic solution qualitatively, but significant
enough to avoid such a prompt outflow. Given that the emissivities are
sensitive functions of temperature and composition, an accurate
expression for these and at least an approximate treatment of neutrino
optical depths is clearly a crucial ingredient in the evolution.  Note
that since the optical depth to neutrinos is never extremely large,
once emitted they are relatively free to escape and are not advected
with the flow.

\subsection{Thermodynamics inside the disk}\label{therm}

\begin{figure} 
  \begin{center}
    \includegraphics[width=0.8\textwidth]{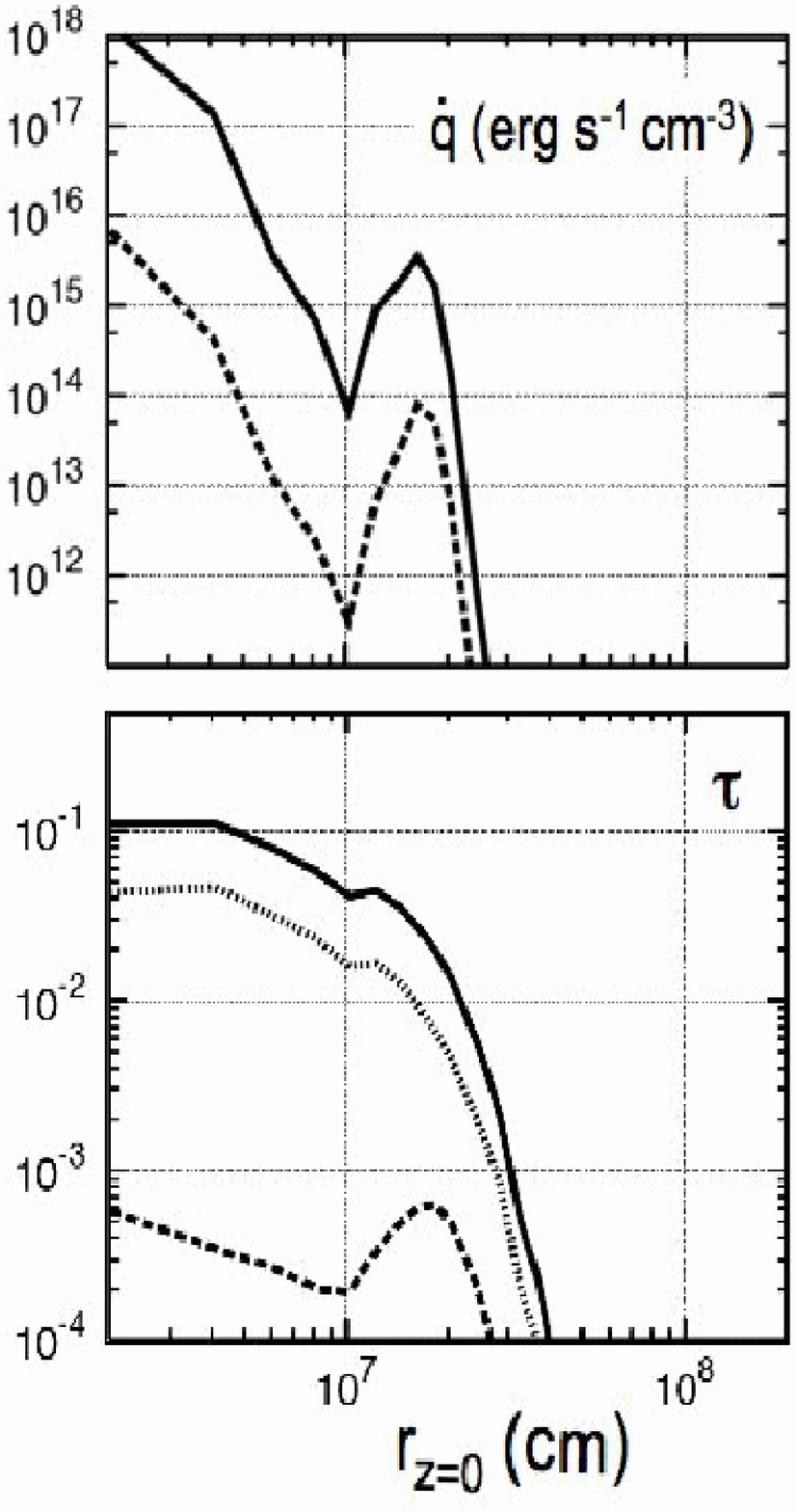} 
    \caption{Upper panel: neutrino emissivities $\dot{q}_{ann}$ (solid
      line) and $\dot{q}_{cap}$ (dashed line) for $(J_0,\alpha)=(3
      r_{\rm g}c,0.1)$ at $t=0.2$~s. Lower panel: the corresponding
      optical depth contributions from $\tau_{\rm abs-ann}$ (solid
      line), $\tau_{\rm abs-cap}$ (dashed line), $\tau_{\rm scat}$
      (dotted line) for $(J_0,\alpha)=(3 r_{\rm g}c,0.1)$ at
      $t=0.2$~s.}
    \label{qdots1}
  \end{center}
\end{figure}

The origin of the energy release and the associated neutrino
luminosity can be easily understood by examining the various emission
mechanisms separately. Figure~\ref{qdots1} shows a snapshot at
$t=0.2$~s of the neutrino emissivities for pair annihilation
($\dot{q}_{ann}$) and $e^{\pm}$ capture onto free nucleons
($\dot{q}_{cap}$) as a function of the cylindrical radius $r$ in the
equatorial plane (z=0), for $(J_0,\alpha)=(3 r_{\rm g}c,0.1)$.

\begin{figure}
  \begin{center}
    \includegraphics[width=0.6\textwidth]{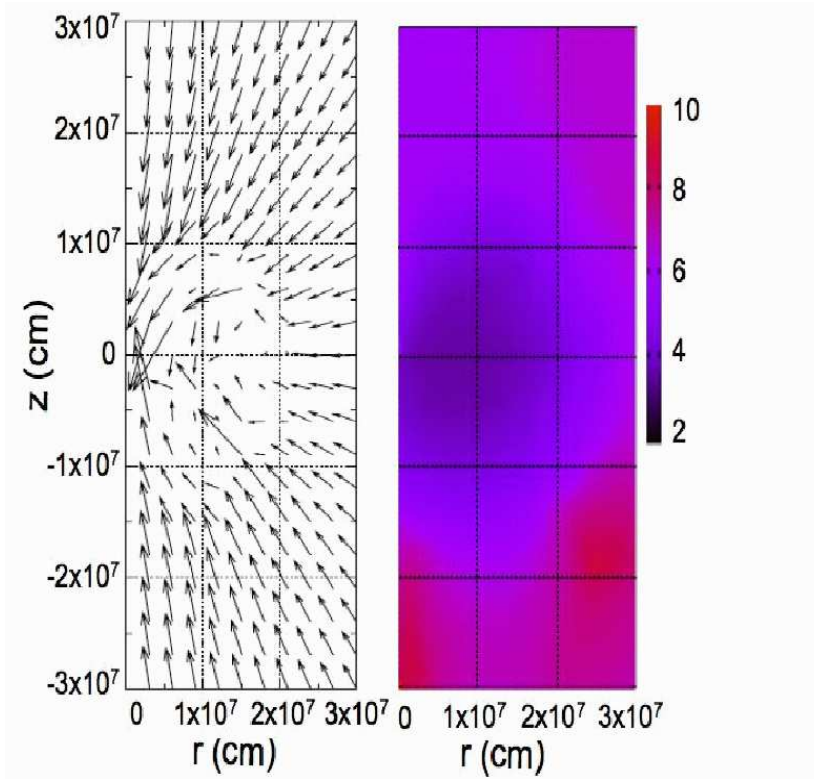}
      \caption{Map of $log_{10}\left(t_{cool} / t_{dyn} \right)$ for $(J_0,\alpha)=(3 r_{\rm
          g}c,0.1)$ at $t=0.2$~s. For the velocity field, the largest vectors are the same as in Figure \ref{vel2}.}
    \label{cooldyn}
  \end{center}
\end{figure}

The highest energy release occurs very close to the BH: at
$r\le 2 \times 10^{7}$~cm both $\dot{q}_{cap}$ and $\dot{q}_{ann}$
rise at least 3 orders of magnitude and release up to
$10^{18}$~erg~s$^{-1}$~cm$^{-3}$.  The figure also shows the
corresponding optical depth contributions for both processes as well
as coherent scattering ($\tau_{\rm scat}$) for $(J_0,\alpha)=(3 r_{\rm
  g}c,0.1)$ at $t=0.2$~s. The inner regions ($r\le 2 \times
10^{7}$~cm) can become somewhat opaque while at large radii the fluid
remains optically thin.  Electron/positron annihilation dominate the
optical depth in the dense inner region over captures onto free
nucleons and coherent scattering. Neutrinos are not entirely free to
escape and corrections due to finite optical depth both in the
pressure and luminosity (Eq. \ref{qnu}) must be taken into account. A
clue to the disk's structure can be inferred from the comparison of
the local cooling (or Kelvin-Helmholtz) time, $t_{\rm cool}\approx
E/\dot{q}_{\nu}$ to the dynamical time, $t_{\rm dyn}\approx (GM_{\rm
  BH}/r^{3})^{-1/2}$, plotted in Figure~\ref{cooldyn}. Despite the
high accretion rates, even in the inner regions the cooling time is
much longer than the orbital period \footnote{At large radii it is
  appropriately even larger, since the stellar envelope is essentially
  in hydrostatic equilibrium.}. This is in agreement with the previous
comparison to the isothermal and adiabatic limits in
Section~\ref{coolingexple}, where we already found that qualitatively at
least, the overall picture is indicative of inefficient cooling.

Regarding the thermodynamical properties of the flow, we will focus on
$(J_0,\alpha)=(3 r_{\rm g}c,0.1)$ and $(J_0,\alpha)=(2 r_{\rm
  g}c,0.1)$ (the remaining cases are quite similar). Equatorial
($z=0$) profiles for density, temperature, electron fraction and
nucleon mass fraction are shown in Figure~\ref{rho4z} for both
runs. The outwardly moving accretion shock front in the high angular
momentum case is visible at $r\approx 2 \times 10^{7}$~cm. Once in the
postshock region, compression and the rise in temperature fully break
up $\alpha$ particles into their constituent neutrons and protons. The
consequent cooling through $e^{\pm}$ captures onto neutrons and
protons allows the gas to reach densities high enough that
neutronization takes place, lowering $Y_{e}$ to minimum values close
to 0.2 (the higher densities also make the fluid become more
degenerate).

For the calculation with $J(R)=J_{\rm crit}$ no centrifugally
supported disk forms, although compression is important in the flow
just before the gas falls through the inner boundary. The equatorial
profiles (shown in the previous figures along with those for
$J_0=3.0r_{\rm g} c$) show little evolution, and this case is basically in
the inviscid limit, where varying $\alpha$ makes little or no
difference on the outcome. The gas is so close to being in free fall
that the much longer viscous timescale becomes irrelevant, and runs with
different values of $\alpha$ yielded the same solution.

Contrary to the cases with $J(R) > J_{crit}$, where the disk was in
some cases opaque, the maximum optical depth is now at most
$\tau=0.01$, dominated also by pair annihilation (the temperature is
$kT\simeq 4$~MeV) and the emissivities are substantially
lower. Compression and the associated rise in temperature are strong
enough to dissociate He almost completely in the inner regions, but no
significant neutronization occurs because the maximum density is only
$\simeq 10^{9}$~g~cm$^{-3}$. Very close to the BH, $e^{\pm}$
captures produce slight neutronization and the electron fraction drops
down to $Y_e=0.45$. The $e^{\pm}$ pairs are at the threshold of
degeneracy, with $\mu_{\rm e}/k_{\rm B}T \approx 1.5$, where $\mu_{\rm
  e}$ is the chemical potential of the electrons.

The possibility of this transition at high accretion rates (of order a
few tenths of a solar mass per second here) was hinted at in the
simpler calculations of \citet{lrr06}, but the equation of state used
there did not permit neutronization or degeneracy effects to play a
role \footnote{They assumed that hot $e^{\pm}$ pairs were abundant,
  producing a pressure contribution $\propto T^{4}$, and that
  $\tau_{\nu} \ll 1$.}.  It is clear from the present set of
calculations that such approximations are valid only for quite lower
accretion rates (a few hundredths of a solar mass per second). This
limitation is valid also for other collapsar studies, where neither
finite degeneracy or neutronization have been fully considered. Two
important potential consequences in this context stand out.

First, the energy of a given neutrino depends on the physical process
responsible for its creation. Those arising from pair annihilation (a
thermal process) have characteristic energies of the order $E_{\rm
  ann}\approx4 kT$, while those due to $e^{\pm}$ capture onto free
nucleons (a weak interaction) have energies of the order $E_{\rm
  cap}\approx E_{\rm F}$, where $E_{\rm F}\simeq 9 (\rho_{10}Y_{\rm
  e})^{1/3}$~MeV is the Fermi energy of electrons/positrons
($\rho_{10}=\rho/10^{10}$~g~cm$^{-3}$).  Table~2 shows the values for
$\rho_{10}$, $T_{10}$, $Y_{e}$, $E_{ann}$, and $E_{cap}$ for our
calculations (in the inner regions where the energy release is
largest, i.e., where $\rho=\rho_{\rm max}$, $T=T_{\rm max}$, and
$Y_{\rm e}=Y_{\rm e, min}$). As $E_{\rm ann}$ and $E_{\rm cap}$ vary
with respect to $J_0$ and $\alpha$, the transition to a different
thermodynamic regime will thus modify the emergent neutrino
spectrum. This in turn affects the global energy release, since at a
given neutrino luminosity, $L_{\nu}$, the efficiency for neutrino
annihilation scales with $<E_{\nu}>$, and thus more energetic
neutrinos will be more efficiently converted into a relativistic pair
plasma.

\begin{table}\centering
\caption{Neutrino Energetics}
\begin{tabular}{ccccccc}
\hline
\hline
$J_{0} $ & $\alpha$ & $\rho_{10}$ & $T_{10}$ & $Y_{e}$ & $E_{\rm ann}$ & $E_{\rm cap}$ \\
($r_{\rm g} \ c$) &  &  & &  & (MeV) & (MeV) \\
\hline
2.0 & 0.10   & 0.07   & 3.70 & 0.44 & 12.76 & 2.87  \\
2.5 & 0.10   & 1.02   & 4.20 & 0.23 & 14.46 & 5.52  \\
3.0 & 0.10   & 0.82   & 3.48 & 0.18 & 12.00 & 4.74  \\
3.0 & 0.01   & 4.00   & 3.75 & 0.10 & 12.94 & 6.63   \\
3.0 & 0.00   & 1.89   & 4.87 & 0.12 & 16.77 & 5.48  \\
\hline
\hline  \\
\end{tabular}
\end{table}

Second, the morphology of the flow may be influenced by the available
energy sinks (e.g., advection, neutrino losses) allowed in a
calculation. Dissipation in the disk through viscosity adds to the
internal energy reservoir. If it is allowed to escape with moderate
efficiency the vertical scale height will be limited. Otherwise the
disk will tend to expand in response, or possibly drive winds from its
surface. In the original collapsar calculation of \citet{mw99}, strong
winds driven from the surface of the disk at small radii were reported
for one calculation, with $\alpha=0.10$. Calculations with a smaller
dissipation rate did not exhibit this feature. For now, our
computations do not span such a long time interval, but the essential
morphological features (expanding outer accretion shock, hot torus, and
dense inner disk) are well established by 0.4~s. We intend to further
explore these issues in a second set of simulations with varying
angular momentum distributions. Any potential outflows emanating from
the inner regions could have a significant impact on the observable
signature of the stellar collapse, and so this deserves careful
consideration.

\subsection{Accretion rate, neutrino luminosity, and energy conversion efficiency.}\label{mdot_eff_lum}

\begin{figure}
  \begin{center}
    \includegraphics[width=0.6\textwidth]{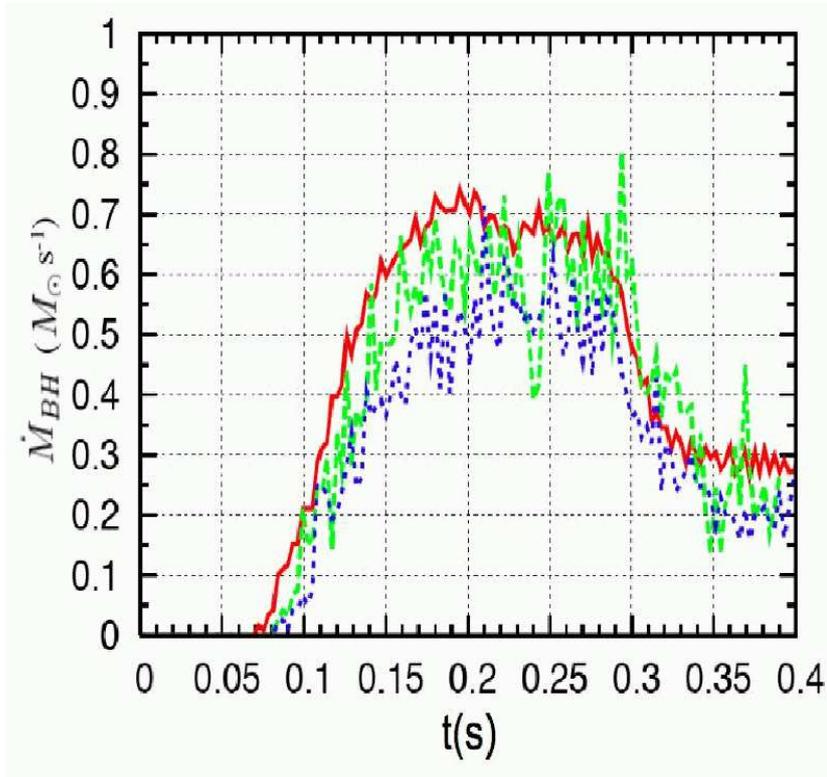}
      \caption{Mass accretion rate onto the BH ($\dot{M}_{\rm
          BH}$) for a fixed $\alpha$ with different angular momentum
        values: $(J_0,\alpha)=(2 r_{\rm g}c,0.1)$ (red solid line),
        $(J_0,\alpha)=(2.5 r_{\rm g}c,0.1)$ (green dashed line), and
        $(J_0,\alpha)=(3 r_{\rm g}c,0.1)$ (blue dotted line).}
    \label{mdot1}
  \end{center}
\end{figure}

\begin{figure}
  \begin{center}
    \includegraphics[width=1.0\textwidth]{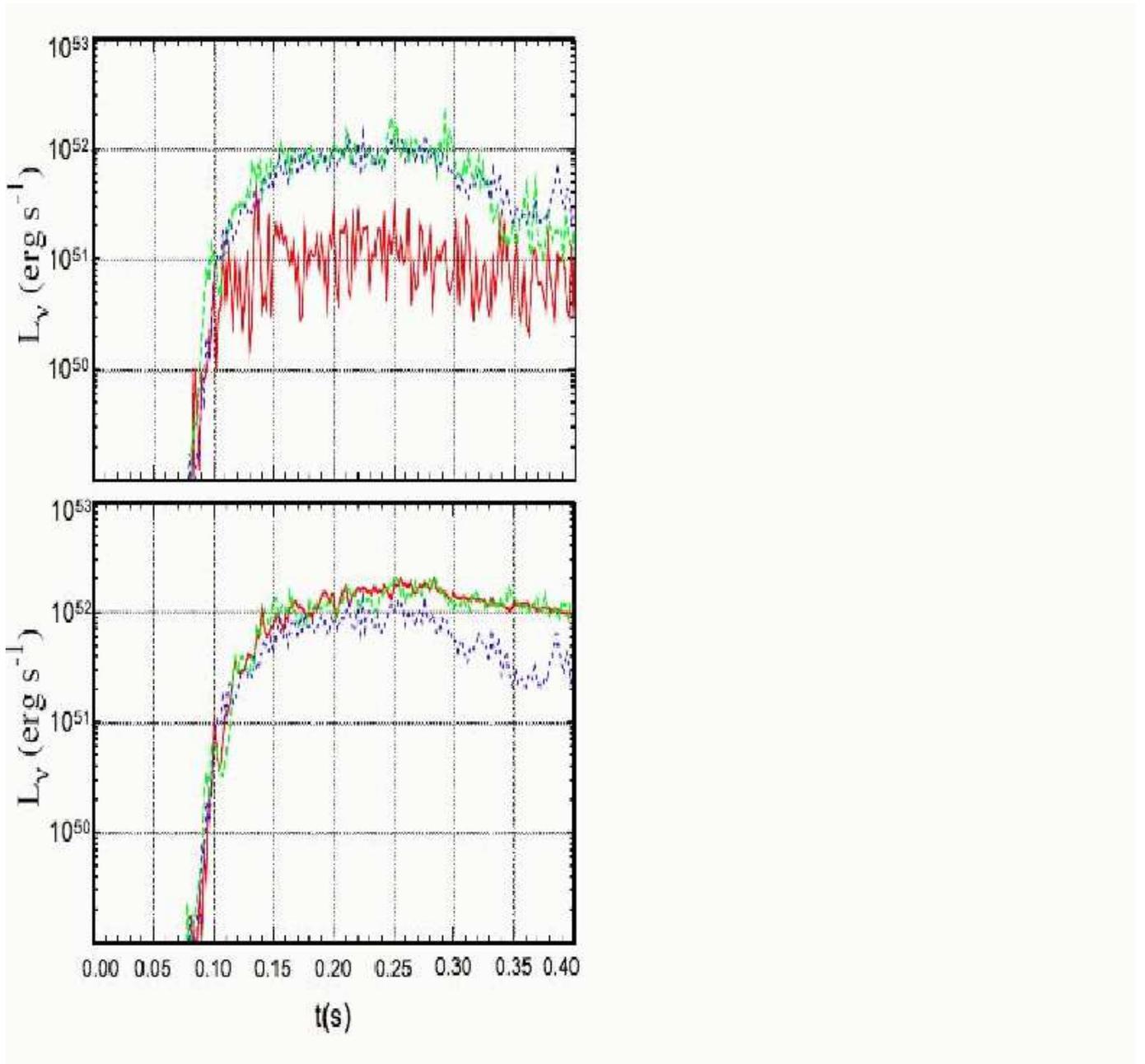} 
      \caption{Upper panel: Neutrino luminosity $L_{\nu}$ for a fixed
        value of $\alpha=0.10$ and different angular momentum values:
        $(J_0,\alpha)=(2 r_{\rm g}c,0.1)$ (red solid line),
        $(J_0,\alpha)=(2.5 r_{\rm g}c,0.1)$ (green dashed line), and
        $(J_0,\alpha)=(3 r_{\rm g}c,0.1)$ (blue dotted line). Lower
        panel: Neutrino luminosity $L_{\nu}$ for a fixed angular
        momentum $J_0=3.0 r_s c$ and varying $\alpha=0.00$ (red solid
        line), $\alpha=0.01$ (green dashed line), and $\alpha=0.10$ (blue
        dotted line). }
    \label{lum1}
  \end{center}
\end{figure}

The mass accretion rate $\dot{M}_{BH}(t)$ was computed at the inner
boundary $R=R_{in}=20$~km and is shown in Figure~\ref{mdot1} for three
runs with the same value of $\alpha=0.10$, and different angular
momentum $J_{0}$. The corresponding neutrino luminosity is plotted in
the top panel of Figure~\ref{lum1}. The initial delay of about 0.1
seconds represents the infall time from the outer boundary of the iron
core. Thereafter the accretion rate and the luminosity rise rapidly,
reaching approximately 0.6 solar masses per second and
$10^{51}-10^{52}$~erg~s$^{-1}$ respectively.  The inner disk is
responsible for most of the energy release, and its formation and
steady configuration for the next 0.1-0.2 seconds produce a nearly
constant accretion rate and luminosity. The efficiency
$L_{\nu}/\dot{M}_{\rm BH} c^{2}$ during this period is approximately
0.01 and 0.001, for high and low angular momentum, respectively,
reflecting the increased importance of advection when the flow is
quasi-radial. Note that despite the absence of a centrifugal barrier
and a dense disk, the configuration with low angular momentum is
capable of significant energy release ($ L_{\nu} \simeq
10^{51}$~erg~s$^{-1}$).

The launching of the accretion shock after a few tenths of a second
leads to a drop in both the accretion rate and the luminosity, while
maintaining a nearly constant efficiency. For $J_{0}/\left(r_{\rm g}
c\right)=2.5, 3.0$ the outwardly propagating shock eventually perturbs
the mass flux to the inner disk before reaching the outer boundary
used in the calculation.  The higher mass accretion rate at low
angular momentum (below the critical value) is due to GR effects but
does not initially translate into a higher luminosity because the
material falls nearly radially into the BH and is unable to
radiate efficiently.

Neutrino luminosities for a fixed value of angular momentum
($J_{0}=3.0r_{\rm g} c$) and varying efficiency of viscous transport are
shown in the bottom panel of Figure~\ref{lum1}. A moderate to null
viscosity is in this early stage the most efficient at producing
neutrinos in spite of the lower dissipation rate since very efficient
angular momentum transport drains the inner disk much too rapidly.

Previous studies \citep{dmpn02,lrrp04,lrrp05} have shown that at very
high accretion rates, greater than approximately one solar mass per
second, the neutrino luminosity saturates and the accretion efficiency
decreases, inhibiting the driving of winds from the disk
\citep{dmpn02}. For the cases computed here, regardless of the actual
value of angular momentum considered, the flow never becomes strongly
opaque to its own neutrino emission in the sense that the emitted
energy is able to escape, thus avoiding this limitation. Resolving
this issue in more detail requires exploring a wider range of
progenitors and longer timescales than those considered here and is
beyond the scope of the present paper. \\

\section{Discussion and Conclusions}\label{sec:conc}

\subsection{Limitations and comparison to other work}

As with all numerical work, the choices made in carrying out the
simulations reflect intentions and biases, and the current
investigation lacks in several aspects. For example, being a
two-dimensional calculation, self-gravity has been considered only in
an approximate manner, and stability issues that relate to this or are
intrinsically three-dimensional (such as spiral arms) are
ignored. 

Magnetic fields have not been included and are in all likelihood
important in several aspects, two of which deserve special
mention. The first is simply that magnetic fields are thought to be at
the origin of the magneto rotational instability (MRI) \citep{bh98}
responsible for angular momentum transport and dissipation, which we
have considered through the $\alpha$ parameter. The MRI likely
operates in collapsars, although the details of how it does so are
likely to be different than what we can infer about it from theory and
observation of accretion disks in CVs and X-ray binaries in a more
leisurely state of affairs. The second point is related to the
importance of magnetic pressure in possible outflows emanating from a
disk. Detailed GR MHD calculations \citep{devilliers05} show that a
flow starting from an equilibrium torus develops the MRI and produces
an inner dense disk, a funnel wall, and a corona around the central
BH. Gas pressure dominates largely in the first, so our
hydrodynamical solution to the inner disk is likely to be a realistic
approximation, while magnetic pressure is relatively more important in
the other two, and being polar structures they are important for the
driving of outflows and energy release in the context of GRBs
\citep{mckinney07}. For very rapidly rotating BHs,
\citet{krolik05} have shown that the magnetic field can actually have
a dynamical effect on the mass accretion rate, suppressing it in the
inner regions (in their case the Kerr parameter was $a=0.998$),
although a full evaluation of this deserves further study.

GR, approximated in this work simply by use of the
Paczy\'{n}ski--Wiita potential, is likely to play a role as well. The
location of the horizon and last stable orbit are functions of the
rotation of the BH, lying at $r_{\rm H}=2GM/c^{2},GM/c^{2}$
and $r_{\rm ISCO}=6GM/c^{2},3GM/c^{2}$ for $a=0,1$ respectively, and
this is important since most of the energy is released at small radii.
For our purposes, however, they are not likely to be crucial since the
relevant radii shift appreciably only for very rapidly rotating holes
(with $a$ above 0.7 or so), and the progenitor stars do not easily
reach such high values (model 16TI has $a=0.44$). A secondary aspect
of GR is related to the efficiency for $\nu\overline\nu$
annihilation and the corresponding energy release. Some energy is
directly lost to the BH because of the strong gravitational
field, while focusing increases the annihilation efficiency. Recently
\citet{birkl07} have carefully computed the energy deposition rates
and efficiencies for annihilation in the Kerr metric for various BH spin rates and flow geometries. They find that the power output
is affected by approximately up to a factor of 2, depending on
whether the disk is geometrically thin or thick (or even for spherical
configurations of the neutrinosphere). Thus the general energy scale
is affected, but not crucially so.

On the other hand, our effort has been directed here at improving the
thermodynamical treatment and that of neutrinos, as well as in
considering realistic initial conditions derived from stellar
evolution calculations. The choice of inner boundary means that the
dense, inner disk producing most of the luminosity is well resolved
and treated appropriately. A final point that deserves improvement is
clearly the distribution of angular momentum, since having a constant
value at all radii is not realistic. We have chosen it for now to
gauge the effect of other changes when compared to previous work, and
provide a guide in further investigation, which will fully explore
situations in which the radial part is a characteristic function of
radius (D. Lopez-Camara et al. 2008, in preparation).

With this in mind, we may consider the implications for GRB production
and GRB/SN associations following core collapse and prompt BH formation from this set of calculations.

\subsection{Global energetics}

The fact that we have placed the inner numerical boundary fairly close
to the BH (at $r_{\rm in}=20$~km) allows us to directly and
more realistically compute the energy release (obviously at the cost
of a shorter simulation in time). This is similar to what was done by
\citet{lrr06} and so we can directly compare the two sets of
results. The two most important differences between the two studies
are in the use of a better equation of state and neutrino treatment
here, and in the use of initial models taken from stellar evolution
calculations. The neutrino luminosities are higher in the present
study, partly because of the more realistic physics employed, but also
because of the initial conditions, which directly result in higher
accretion rates and densities.  Since the cooling rate from captures
scales with the density, this fact alone will raise the
luminosity. The rates are quite sensitive to the temperature, but the
rise in the postshock material is mostly due to the infall kinetic
energy per unit mass and is fixed by the potential well of the central
BH.  It is clear as well, from the results presented for
simplified cases in the adiabatic and isothermal limits, and from the
computation of the Kelvin--Helmholtz timescale, that correctly
accounting for the energy sinks is a crucial ingredient if one wishes
to estimate the global energy release and flow properties.  Finally,
the emergent neutrino spectrum is a function of the angular momentum
of the infalling gas, through the selection of the dominant cooling
mechanism. This is admittedly a difficult issue to resolve
observationally, but may impact upon the energy deposition rate.

The two most important variables that determine the global morphology
are the rotation rate, quantified here through the distribution of
specific angular momentum, and the strength of angular momentum
transport, parameterized through the prescription of Shakura \&
Sunyaev. It is interesting to note that for high angular momentum
(larger than the threshold for disk formation), viscous dissipation is
responsible for the production of large-scale flows, namely by
producing an outward moving shock after a certain delay (a few
tenths of a second). The effect is to perturb the flow into the inner
disk and decrease the luminosity. The impact on the flow at large
radii remains to be explored in greater datail due to the limitation of our outer
boundary.

\subsection{Implications for GRB production}

The compression for low angular momentum cases releases a large amount
of energy, not more than 1 order of magnitude smaller than for the
case with a disk, and we believe this to be an important issue: slowly
rotating models are in principle capable of releasing an amount of
energy that could produce GRBs, although admittedly perhaps only at
the faint end of the luminosity distribution. For a 1\% efficiency of
conversion into pairs through $\nu\overline{\nu}$ annihilation at
$L_{\nu}=10^{53}$~erg~s$^{-1}$, the models shown here can produce
annihilation luminosities of about $10^{48}-10^{49}$~erg~s$^{-1}$, depending on
the value of $J_{0}$. Performing realistic runs for longer times is
a priority, but clearly over long timescales, having slow
rotation is not necessarily a handicap regarding the energy
release. The simple reason is that, as known for a long time, and
first quantified in this scenario by \citet{lrr06}, it is most
efficient in terms of energy extraction to have the material release
its energy as close as possible to the BH, or equivalentely,
as deep as possible in the potential well, but not too close so that
it is still shocked by the presence of a centrifugal barrier.
Magnetic mechanisms may also power fully or partially a potential
burst, and we note here only that the internal energies are high
enough, as in the most common hypercritical accretion scenarios, to do
so if a fraction of this is transferred to a magnetic field (at a
level of 10\% of equipartition).

The obvious implication in terms of the type of progenitor one can
consider for GRBs is that single stars at the threshold for disk
formation through centrifugal support are capable of giving a GRB. Having a substantial amount of rotation will obviously
guarantee an accretion disk, but perhaps not all cases require such
special conditions (as for example, torquing of the pre-SN star
by a binary companion). The isotropic equivalent energy of GRB060505
is $\log[E_{\gamma, \rm iso}]\simeq 49.5$, clearly within the range
obtainable with the models presented here even at low rotation rates.

\subsection{Nucleosynthesis in the outflow and GRB/SN association}

Centrifugally supported collapsar disks are expected to produce strong
winds \citep{mw99}, driven quite generically by a combination of
viscous, neutrino, and magnetic effects (the first two of which we have
explicitly included here). \citet{pth04} in particular, computed the
expected nucleosynthesis of $^{56}$Ni in these outflows, using the
steady state disk models of \citet{pwf99} as an initial condition and
considering various mass accretion and dissipation rates. The
essential feature of these models is the computation of the change in
entropy per baryon, $s_{b}/k$ in the outflow, starting from the
midplane of the accretion disk and reaching a large distance at which
the velocity approaches an asymptotic value $v_{\rm w}/c \simeq
0.1-0.2$. From our hydrodynamical determination of the disk structure
using an improved equation of state and thermodynamics, we computed
the equivalent entropies in the disk and applied the formalism of
\citet{pth04} to compute the total change in $s_{b}/k$ (assuming the
wind reaches similar velocities, unresolved in our simulations
spanning 0.5~s). The mass fraction converted to $^{56}$Ni, $X_{\rm
  Ni}$, then depends sensitively on $s_{b}/k$ and the ratio
$\beta=\dot{M}_{\rm o}/v_{\rm w}^{3}$, where $\dot{M}_{\rm o}$ is the
mass outflow rate and $v_{\rm w}$ is the terminal wind velocity. Once
the outflow is launched we can obtain an estimate of $\dot{M}_{\rm o}$
directly from the momentum field in the simulation, and $v_{\rm w}/c$
is taken as $\simeq 0.2$. We find that for $J_{0} > J_{\rm crit}$, and
independently of the strength of angular momentum transport,
substantial Ni synthesis occurs (X$(^{56}\mbox{Ni})\approx 0.5$) in an
outflow with $\dot{M}\approx 0.3$~M$_{\odot}$~s$^{-1}$. This would
imply a total ejection of $\simeq M_{\odot}$ in $^{56}$Ni in a GRB
lasting 10~seconds, in the right range to account for the outflows
seen, e.g., in SN2003dh. The relatively narrow range in $J_{0}$
spanned in our calculations does not allow for a good extrapolation to
higher rotation rates, but clearly the amount of mass involved is
significant, and would likely grow substantially at higher $J_{0}$
where more masive and more radially extended disks are expected
(\citet{pth04} estimate several $M_{\odot}$ for the massive,
centrifugally supported disks taken from \citet{pwf99}). A
configuration in which no significant outflows occur because the
rotation rate is too low ($J \le J_{\rm crit}$) would not be expected
in this scenario to produce any significant ejection of radioactive
elements capable of producing a SN-like signature. It could, as shown
here, still release enough energy through neutrinos or magnetic
mechanisms to power a classical GRB.  The result, if it were to occur
at low redshift, could possibly be an event resembling GRB060505.

\acknowledgments 
We thank S.E. Woosley and A. Heger for making their
pre-SN models available. This work was supported in part by
CONACyT (45845E, WL), PAPIIT-UNAM (IN113007, WL), NASA (Swift
NX07AE98G, ER-R) and DOE SciDAC (DE-FC02-01ER41176, ER-R). D.L.-C.
acknowledges support through a CONACyT graduate scholarship. W.H.L.
thanks the Department of Astronomy and Astrophysics at the University
of California, Santa Cruz for hospitality. We thank an anonymous
referee for his comments and suggestions in improving the paper.


\begin{thebibliography}

\bibitem[Balbus \& Hawley(1998)]{bh98}Balbus, S. A. \& Hawley,
  J. F. 1998 Rev. Mod. Phys., 70,. 1
\bibitem[Beloborodov(2003)]{b03} Beloborodov, A. M. 2003, \apj, 588, 931
\bibitem[Beloborodov \& Illarionov(2001)]{bel01}Beloborodov, A. M., \& Illarionov, A. F. 2001, \mnras, 323, 167
\bibitem[Birkl, Aloy \& Janka(2007)]{birkl07}Birkl, R., Aloy, M. A. \& Janka, H. T., 2007, A\&A, 463, 51
\bibitem[Blinnikov, Dunina-Barkovskaya \& Nadyozhin(1996)]{bdbn96} Blinnikov, S. I., Dunina-Barkovskaya, N. V., \& Nadyozhin, D. K. 1996, \apj, 106, 171
\bibitem[Bloom, Kulkarni \& Djorgovski(2002)]{bkd02} Bloom, J. S., Kulkarni, S. R., \& Djorgovski, S. G. 2002, \apj, 123, 1111
\bibitem[Campana et al.(2006)]{c06} Campana, S., et al. 2006, Nature, 442, 1008
\bibitem[Cantiello et al.(2007)]{cantiello07}Cantiello, M.; Yoon,
S.-C.; Langer, N.; Livio, M. 465, L29
\bibitem[Chevalier(1989)]{c89} Chevalier, R.A. 1989, \apj, 346, 847
\bibitem[Della Valle et al.(2003)]{dv03} Della Valle, M., et al. 2003, \aap, 406, L33
\bibitem[Della Valle et al.(2006a)]{dv06apj} Della Valle, M., et al. 2006a, \apjl, 642, L103
\bibitem[Della Valle et al.(2006b)]{dv06nat} Della Valle, M., et al. 2006b, Nature, 444, 1050
\bibitem[Dessart et al.(2008)]{dessart08}Dessart, L., Burrows, A.,
Livne, E., Ott, C. D. 2008, \apj, 673, L43
\bibitem[Detmers et al.(2008)]{detmers08}Detmers, R. G.; Langer, N.;
Podsiadlowski, Ph.; Izzard, R. G. 2008 A\&A, 484, 831
\bibitem[Dezalay et al.(1996)]{d96} Dezalay, J. P., et al. 1996, \apj, 471, L27
\bibitem[De Villiers, Hawley \& Krolik(2005)]{devilliers05}De Villiers, J.-P., Hawley, J. F. \& Krolik, J. H. 2005, \apj, 599, 1238
\bibitem[Di Matteo, Perna \& Narayan(2002)]{dmpn02} Di Matteo, T., Perna, T., \& Narayan, R. 2002, \apj, 579, 706
\bibitem[Eichler et al.(1989)]{e89}Eichler, D., Livio, M., Piran, T., Schramm, D.N. 1989, Nature, 340, 126
\bibitem[Fishman \& Meegan(1995)]{fm95}Fishman, G. J., Meegan, C. A. 1995, ARA\&A, 33, 415
\bibitem[Fynbo et al.(2006)]{f06} Fynbo, J. P. U., et al. 2006, Nature, 444, 1047
\bibitem[Fruchter et al.(2006)]{fruchter06}Fruchter, A. S. et al. 2006, Nature, 441, 463
\bibitem[Galama et al.(1998)]{g98} Galama, T. J., et al. 1998, Nature, 395, 670 
\bibitem[Gal Yam et al.(2006)]{gy06} Gal Yam, A., et al. 2006, Nature, 444, 1053
\bibitem[Gehrels et al.(2007)]{gcn07}Gehrels, N., Cannizzo, J.K., Norris, J.P. 2007, New J. Phys., 9, 37
\bibitem[Gehrels et al.(2006)]{g06} Gehrels, N., et al. 2006, Nature, 444, 1044
\bibitem[Gorosabel et al.(2005)]{g05} Gorosabel, et al. 2005, \aap, 444, 711
\bibitem[Heger et al.(2005)]{hws05}Heger, A., Woosley, S.E., Spruit, H.C. 2005, \apj, 626, 350
\bibitem[Heger et al.(2000)]{h00} Heger, A., et al. 2000, \apj, 528, 368
\bibitem[Houck \& Chevalier(1991)]{hc91} Houck, J.C., Chevalier, R.A., 1991, \apj, 376, 234
\bibitem[Itoh et al.(1996)]{i96} Itoh, N., et al. 1996, ApJS, 102, 411
Itoh et al. (1996)
\bibitem[Izzard, Ramirez-Ruiz \& Tout(2004)]{izzard04}Izzard, R.G.,
Ramirez-Ruiz, E., Tout, C. A. 2004 MNRAS, 348, 1215

\bibitem[Janiuk \& Proga(2008)]{jp08} Janiuk, A., \& Proga, D. 2008, \apj, 675, 519
\bibitem[Janiuk, Yuan, Perna \& Di Matteo(2007)]{jypdm07} Janiuk, A., Yuan, Y., Perna, R., \& Di Matteo, T. 2007, \apj, 664, 1011
\bibitem[Kaneko et al.(2008)]{kaneko08}Kaneko, Y., et al. 2008, \apj,
654, 385
\bibitem[Klebesadel et al.(1973)]{kso73}Klebesadel, R.W., Strong, I.B., Olsson, R.A. 1973, \apj, 182, L85 
\bibitem[Krolik, Hawley \& Hirose(2005)]{krolik05}Krolik, J. H., Hawley, J. F. \& Hirose, S. 2005, \apj, 622, 1008
\bibitem[Lattimer \& Schramm(1974)]{ls74} Lattimer, J. M., \& Schramm, D. N. 1974, \apj, 192, L145
\bibitem[Lattimer \& Schramm(1976)]{ls76} Lattimer, J. M., \& Schramm, D. N. 1976, \apj, 210, 549
\bibitem[Langanke \& Mart\'{\i}nez-Pinedo(2001)]{lmp01} Langanke, K., \& Martt\'{\i}nez-Pinedo Lattimer, G. 2001, At. Data Nucl Data Tables, 79, 1
\bibitem[Lee \& Ramirez-Ruiz(2002)]{lrr02} Lee, W. H., \& Ramirez-Ruiz, E. 2002, \apj, 577, 893
\bibitem[Lee \& Ramirez-Ruiz(2006)]{lrr06} Lee, W. H., \& Ramirez-Ruiz, E. 2006, \apj, 641, 961
\bibitem[Lee \& Ramirez-Ruiz(2007)]{lrr07} Lee, W. H., \& Ramirez-Ruiz, E. 2007, New J. Phys., 9, 17
\bibitem[Lee, Ramirez-Ruiz \& Page(2004)]{lrrp04}Lee, W. H., Ramirez-Ruiz, E. \& Page, D. 2004, \apjl, 608, L5 
\bibitem[Lee, Ramirez-Ruiz \& Page(2005)]{lrrp05} Lee, W. H., Ramirez-Ruiz, \& E., Page, D. 2005, \apj, 632, 421
\bibitem[MacFadyen \& Woosley(1999)]{mw99} MacFadyen, A. I., \& Woosley, S. E. 1999, \apj, 524, 262
\bibitem[McKinney \& Narayan(2007)]{mckinney07}McKinney, J. C. \& Narayan, R. 2007, \mnras, 375, 513
\bibitem[Malesani et al.(2004)]{m04} Malesani, D., et al. 2004, \apj, 609, L5
\bibitem[Meegan et al.(1992)]{m92} Meegan, C. A., et al. 1992, Nature, 355, 143
\bibitem[Mendoza et al.(2009)]{mtn09} Mendoza, S., Tejeda, E., \& Nagel, E.\ 2009, MNRAS, 393, 579 
\bibitem[M\'{e}sz\'{a}ros(2002)]{m02}M\'{e}sz\'{a}ros, P. 2002, ARA\&A, 40, 137
\bibitem[M\'{e}sz\'{a}ros \& Rees(1997)]{mr97} M\'{e}sz\'{a}ros, P., \& Rees, M. J. 1997, \apj, 482, L29
\bibitem[Metzger et al.(1997)]{m97} Metzger, M. R., et al. 1997, Nature, 387, 878
\bibitem[Monaghan(1992)]{mon92}Monaghan, J.J. 1992, \araa, 30, 543 
\bibitem[Nakar(2007)]{n07}Nakar, E., 2007, Phys. Rep., 442, 166
\bibitem[Narayan et al.(1992)]{n92}Narayan, R., Paczy\'{n}ski, B., Piran, T. 1992, \apj, 395, L83
\bibitem[Narayan, Piran \& Kumar(2001)]{npk01} Narayan, R., Piran, T., \& Kumar, P. 2001, \apj, 557, 949
\bibitem[Paczy\'{n}ski(1986)]{bp86} Pacz\'{n}yski, B. 1986, \apj, 308, L43
\bibitem[Paczy\'{n}ski(1991)]{bp91} Pacz\'{n}yski, B. 1991, Acta Astron., 41, 257
\bibitem[Paczy\'{n}ski \& Wiita(1980)]{pw80} Pacz\'{n}yski, B., \&
  Wiita, P. J. 1980, \aap, 88, 23
\bibitem[Piran(2004)]{piran04}Piran, T. 2004, Rev. Mod. Phys, 76, 1143
\bibitem[Popham \& Narayan(1995)]{pn95} Popham, R., \& Narayan, R. 1995, \apj, 442, 337
\bibitem[Popham, Woosley \& Fryer(1999)]{pwf99} Popham, R., Woosley, S. E., \& Fryer, C. 1999, \apj, 518, 356
\bibitem[Prochaska et al.(2004)]{p04} Prochaska, J. X., et al. 2004, \apj, 611, 200
\bibitem[Proga \& Begelman(2003)]{pb03} Proga, D., \& Begelman M. C. 2003, \apj, 582, 69
\bibitem[Proga, MacFadyen, Armitage \& Begelman(2003)]{pmab03} Proga, D., MacFadyen, A. I., Armitage, P. J., \& Begelman, M. C. 2003, \apj, 599, L5
\bibitem[Pruet et al.(2004)]{pth04}Pruet, J., Thompson, T. A., Hoffman, R. D. 2004, \apj, 606, 1006
\bibitem[Ramirez-Ruiz et al.(2005)]{rr05}Ramirez-Ruiz, E., Granot, J.,
Kouveliotou, C., Woosley, S.E., Patel, S.K., Mazzali, P.A. 2005, \apj,
625, L91
\bibitem[Salpeter(1964)]{s64}Salpeter, E. E. 1964, \apj, 140, 796
\bibitem[Shakura \& Sunyaev(1973)]{ss73} Shakura, N. I., \& Sunyaev, R. A. 1973, \aap, 24, 337
\bibitem[Shapiro \& Teukolsky(1983)]{st83} Shapiro S. L., \& Teukolsky, S. A. 1983, ``Black Holes, White Dwarfs, and Neutron Stars: The Physics of Compact Stars (New York: Wiley-Interscience).
\bibitem[Soderberg et al.(2005)]{sod05} Soderberg, A. M., et al. 2005, \apj, 627, 877
\bibitem[Sollerman et al.(2005)]{sol05} Sollerman, J., et al. 2005, New Astronomy, 11, 103-115
\bibitem[Spruit(2002)]{spruit02} Spruit, H.C. 2002, A\&A, 381, 923
\bibitem[Stanek et al.(2003)]{s03} Stanek, K. Z., et al. 2003, \apj, 591, L17
\bibitem[Thompson(1994)]{t94} Thompson, C. 1994, MNRAS, 270, 480
\bibitem[Usov(1992)]{u92} Usov, V. V. 1992, Nature, 357, 472 
\bibitem[van Paradijs et al.(2000)]{vkw00}van Paradijs, J., Kouveliotou, C., Wijers, R.A.M.J. 2000, ARA\&A, 38, 379
\bibitem[Woosley(1993)]{w93} Woosley, S. E. 1993, \apj, 405, 273
\bibitem[Woosley \& Bloom(2006)]{wb06}Woosley, S.E., Bloom, J.S. 2006, ARA\&A, 44, 507
\bibitem[Woosley  \& Heger(2006)]{wh06} Woosley S. E., \& Heger A. 2006, \apj, 637, 914
\bibitem[Yoon \& Langer(2005)]{yoon05} Yoon, S.C., Langer, N., 2005, A\&A, 443, 643
\bibitem[Zel'Dovich(1964)]{z64} Zel'Dovich, Y.~B.\ 1964, Soviet Physics Doklady, 9, 195
\end{thebibliography}
\end{document}